\documentclass[journal]{IEEEtran}
 
 \usepackage{amsmath}
\usepackage{amssymb}
\usepackage{graphicx}
\usepackage{booktabs}
\usepackage{algpseudocode}
\usepackage{algorithm}
\usepackage{xcolor}
\usepackage[normalem]{ulem}
\usepackage[most]{tcolorbox}
\usepackage{tabularx}
\usepackage{diagbox}
\newtcolorbox{promptbox}[2][]{%
  enhanced,
  colback=gray!2,
  colframe=black!55,
  colbacktitle=gray!12,
  coltitle=black,
  boxrule=0.45pt,
  arc=0.8mm,
  left=1.5mm,
  right=1.5mm,
  top=1.0mm,
  bottom=1.0mm,
  boxsep=0.4mm,
  fonttitle=\bfseries\footnotesize,
  title={#2},
  #1
}

\DeclareMathOperator*{\argmin}{arg\,min}

\begin{document}
 \bstctlcite{BSTcontrol}
\title{SiteEvo: An LLM Agent for Autonomous Discovery of Site-Adaptive Massive MIMO Detectors}
 
\author{Jiajia~Guo,~\IEEEmembership{Member,~IEEE,} 
    Chao-Kai~Wen,~\IEEEmembership{Fellow,~IEEE,} Shi~Jin,~\IEEEmembership{Fellow,~IEEE,} and Jun~Zhang,~\IEEEmembership{Fellow,~IEEE}  
\vspace{-7mm}
}

\maketitle

\begin{abstract}
Designing a massive multiple-input multiple-output (MIMO)
detector that is tailored to each specific deployment site can improve detection
performance, yet most existing detectors, including deep-unfolding methods, adapt mainly within human-prescribed algorithmic structures.
This letter proposes SiteEvo, a site-adaptive agentic discovery framework that extends adaptation from parameter tuning to detector-algorithm discovery. Rather
than selecting or recombining predefined architectures, the large language model (LLM) agent synthesizes complete executable detector programs under a common design contract by combining its pretrained communication knowledge,
site-specific wireless evidence, and experience accumulated during the
search. Each proposal defines its own processing structure and trainable
parameterization, and the parameters are optimized by a
communication-system backend. The resulting
high-performance detector is further structurally refined to reduce
block-level computational cost under a bit error
rate (BER)-preservation constraint and
is finally frozen for deployment without online LLM inference.
Simulations across three sites demonstrate BER reductions by factors of 2.6 to 6.0 over the best-performing competing detector at each site at 6~dB, while structural refinement reduces block-level computational cost by 16.7-34.1\% under a validation-BER constraint. 
\end{abstract}

\begin{IEEEkeywords}
MIMO detection, agents, algorithm discovery.
\end{IEEEkeywords}
 
\vspace{-3mm}
\section{Introduction}
\IEEEPARstart{M}{assive} multiple-input multiple-output (MIMO)
detection is essential for exploiting the spatial degrees of
freedom of large antenna arrays. In practical deployments,
propagation environments and user distributions vary across
sites, yielding distinct channel statistics and detection
challenges. A detector designed for one channel distribution may not perform equally well at other sites. Such heterogeneity motivates detector designs tailored to site-specific channel distributions, as opposed to a uniform design.
Data-driven MIMO detection includes fully learned methods
that directly learn detection mappings and model-driven
methods that integrate communication priors with trainable
parameters \cite{Shlezinger2022ModelBased}. Deep unfolding
optimizes iterative detection algorithms through trainable
parameters, while recent
mixture-of-experts methods adapt to heterogeneous channels
by selecting among predefined experts \cite{Ren2025MoE}.
Nevertheless, these methods rely on human-prescribed
algorithmic structures or predefined experts, largely
restricting adaptation to parameter learning or model
selection. Manually designing and repeatedly updating
dedicated detector structures for each site is difficult
to scale.
This motivates the automated discovery of detector
structures tailored to site-specific channel characteristics.

Addressing this structure-level adaptation requires a mechanism capable of exploring, evaluating, and refining algorithmic structures. Recent advances in large language models (LLMs) enable autonomous agents that combine reasoning, action, tool use, and feedback. ReAct and Reflexion show that LLM agents can interact with external environments and improve decisions based on execution outcomes \cite{Yao2023ReAct,Shinn2023Reflexion}. Agentic capabilities have also been extended to autonomous scientific discovery, where LLM agents integrate planning, information retrieval, code execution, and experimental tools to explore candidate solutions \cite{Boiko2023Autonomous}. In wireless communications, the AI Telco Engineer (AITE) explored autonomous algorithm discovery for channel estimation and link adaptation using GPT-5.4 and GPT-OSS 120B in separate experiments \cite{Aoudia2026AITEToward}. Subsequent work applied AITE to equalization and pilotless receiver design \cite{Aoudia2026AITE}. Although AITE demonstrates feedback-driven algorithm discovery, its search remains task-oriented over site-adaptive, without explicitly tailoring algorithmic structures to heterogeneous site conditions. How site-specific channel statistics guide the joint discovery of massive MIMO detector structures and trainable parameterizations remains unexplored.

To address this question, we propose \emph{SiteEvo}, a site-adaptive
agentic framework elevating detector adaptation from parameter tuning to
executable algorithm discovery. For each site, the LLM agent combines
pretrained communication knowledge, the system model, channel evidence, and
search experience to synthesize executable detector programs with
trainable parameterizations. A communication-system backend learns their
numerical parameters and feeds validation performance back to subsequent
reasoning, closing the loop among algorithm synthesis, parameter learning, and physical-layer evaluation. The detector is structurally refined under a bit error rate (BER)-preservation constraint and frozen for deployment. Simulations on
a massive MIMO system demonstrate site-specialized
structures with superior performance.

\begin{figure*}[!t]
    \centering
    \includegraphics[width=1.95\columnwidth]{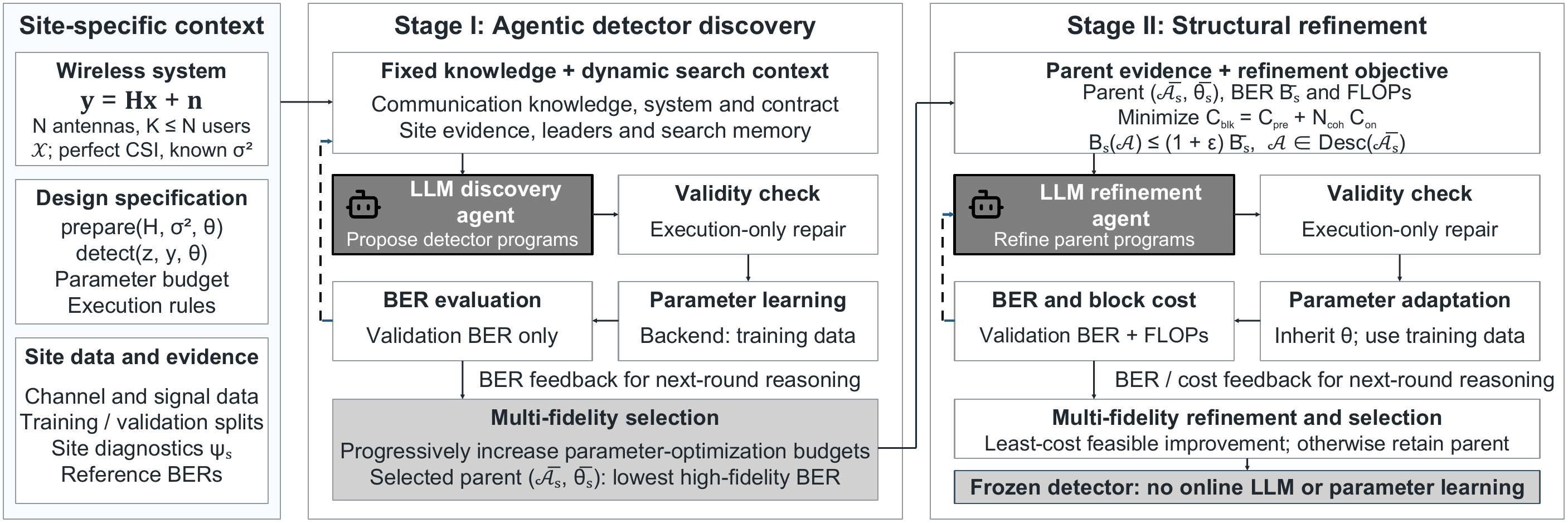}
    \vspace{-3mm}
 \caption{Overall workflow of SiteEvo. Site-specific information, channel evidence, and reference performance guide Stage~I, where the LLM
agent discovers executable detector structures and the communication backend trains parameters and returns validation-BER feedback. Stage~II refines the selected detector for lower block-level complexity under a BER-preservation constraint, yielding a frozen site-specific detector for deployment without online LLM inference.}
    \label{fig:siteevo_overview}
    \vspace{-4mm}
\end{figure*}

\vspace{-4mm} 
\section{System Model}
\label{s2}

We consider an uplink massive MIMO system, where an $N$-antenna base station (BS)
serves $K\leq N$ single-antenna user equipment (UE). The received
baseband signal is
\begin{equation}
\mathbf y=\mathbf H\mathbf x+\mathbf n,
\label{eq:system_model}
\end{equation}
where $\mathbf y\in\mathbb C^N$,
$\mathbf H\in\mathbb C^{N\times K}$, and
$\mathbf x\in\mathcal X^K$ denote the received signal,
channel matrix, and transmitted symbol vector,
respectively, with constellation $\mathcal X$ and
$\mathbb E[|x_k|^2]=1$.
The noise follows
$\mathbf n\sim\mathcal{CN}(\mathbf 0,\sigma^2\mathbf I_N)$.
We assume perfect channel information and known noise variance $\sigma^2$ at the BS.

For the MIMO detection problem, given $\mathbf{y}$, $\mathbf{H}$, and $\sigma^2$, the BS aims to
recover $\mathbf{x}$. As a representative model-driven detector,
OAMP-Net unfolds the orthogonal approximate message passing (OAMP) algorithm into $T$ layers while learning
layer-wise parameters within its prescribed linear-estimation and
denoising structure \cite{He2018ModelDriven}. Hence, data adaptation
changes the numerical parameters but not the underlying detector
structure.
More generally, we represent a detector as $\hat{\mathbf{x}}
    =
    \mathcal{Q}_{\mathcal{X}}
    \left[
        f_{\mathcal{A}}
        \left(
            \mathbf{y},
            \mathbf{H},
            \sigma^2;
            \boldsymbol{\theta}
        \right)
    \right]
    $, where $\mathcal{A}$ denotes the algorithmic structure,
$\boldsymbol{\theta}$ its trainable parameters, and 
$\mathcal{Q}_{\mathcal{X}}[\cdot]$ the constellation decision.
Conventional model-driven learning  prescribes
$\mathcal{A}$ and adapts $\boldsymbol{\theta}$ from data. This
motivates jointly adapting both the detector structure and its
parameters to site-specific channel distributions.

\vspace{-2.5mm}
\section{SiteEvo: Site-Adaptive Agentic Discovery}
\label{sec:siteevo}

To address site-dependent detector design, this section develops SiteEvo,
an agentic framework that jointly adapts detector structures and trainable
parameters. We first formulate the joint design problem and then establish
a two-stage framework comprising performance-oriented detector discovery
and complexity-aware structural refinement.
 
 \vspace{-5mm}

\subsection{Problem Formulation and Framework}
\label{subsec:framework}

Following Section~\ref{s2}, a detector is characterized by two design
degrees of freedom: its algorithmic structure $\mathcal{A}$ and the
associated trainable parameters
$\boldsymbol{\theta}\in\Theta(\mathcal{A})$. Conventional model-driven
learning typically prescribes $\mathcal{A}$ and adapts only
$\boldsymbol{\theta}$ \cite{Shlezinger2022ModelBased}. SiteEvo extends
this paradigm by making the detector structure itself site dependent,
thereby allowing the processing mechanism and its trainable parameters
to be jointly specialized to the statistical characteristics of the
target site.

Let $s\in\mathcal{S}$ index a site and let $\mathcal{D}_{s}$ denote
the channel and signal samples collected from that site. We formulate
the resulting detector design problem as
\begin{equation}
(\mathcal{A}_{s}^{\star},\boldsymbol{\theta}_{s}^{\star})
=
\argmin_{ \mathcal{A}\in\mathbb{A},\, 
\boldsymbol{\theta}\in\Theta(\mathcal{A})}
J_s(\mathcal{A},\boldsymbol{\theta};\mathcal{D}_{s}),
\label{eq:joint_design}
\end{equation}
where $\mathbb{A}$ denotes the set of executable detector structures
satisfying the prescribed design contract, and $J_s(\cdot)$ denotes
the site-specific detection objective, with the BER serving as the principal performance metric in this work. Different site distributions may
therefore favor different processing structures, parameter
configurations, or both. This differs fundamentally from conventional
parameter adaptation, where the processing structure
remains invariant across sites.

Fig.~\ref{fig:siteevo_overview} illustrates the two-stage offline framework. Stage I discovers high-performance detector programs guided by validation BER, while Stage II refines the selected detector to reduce computational cost under a BER-preservation constraint. This separation avoids imposing complexity-driven objectives during initial structural discovery.
 
The site-specific design context comprises three components: (i) the wireless system model; (ii) an executable contract defining \(\operatorname{prepare}(\mathbf H,\sigma^2,\boldsymbol\theta)\) and \(\operatorname{detect}(\mathbf z,\mathbf y,\boldsymbol\theta)\), where \(\mathbf z\) is the reusable channel-dependent state; and (iii) site data \(\mathcal D_s\), diagnostics \(\boldsymbol\psi_s\), and reference BERs. The $\operatorname{prepare}(\cdot)$ function computes the reusable channel-dependent state $\mathbf z$, while $\operatorname{detect}(\cdot)$ uses this state and the received signal $\mathbf y$ to estimate
the transmitted symbols.
The contract imposes execution rules and a trainable-parameter budget without prescribing a detector grammar. The dataset supports parameter learning and validation, while the diagnostics characterize channel correlation, diagonal dominance, and received-power spread. The selected detector \((\overline{\mathcal A}_s,\overline{\boldsymbol\theta}_s)\) becomes the Stage-II parent.
 
\vspace{-3mm}

\subsection{Stage I: Feedback-Guided Agentic Detector Discovery}
\label{subsec:stage1}

The key idea of Stage~I is to discover high-performance detectors
tailored to the target site through an iterative design--evaluation
process. At each search round, the LLM agent draws on its pretrained
communication knowledge, site-specific context, and accumulated search
experience to formulate signal-processing hypotheses and instantiate
them as complete executable detector programs. The agent determines the
processing structure and trainable parameterization, while the
communication-system backend learns the numerical parameters through
gradient-based optimization, as in deep unfolding, and evaluates the
resulting detector. The evaluation results are then returned to guide
subsequent algorithmic reasoning, enabling detector discovery beyond a
predefined detector grammar.
This discovery process proceeds through
four coupled steps, as detailed below.

{\bf (i) Detector-structure generation.} The first task of Stage~I is to
instantiate the structural design variable $\mathcal{A}$ introduced in
\eqref{eq:joint_design}. In contrast to selecting $\mathcal{A}$ from a
predefined detector family, the agent constructs it as an executable
program together with its trainable parameterization
$\Theta(\mathcal{A})$. This structural generation is guided by two
key sources of information: fixed communication knowledge that
defines the detector-design space, and evolving site-specific evidence
that reveals which structures are effective for the target site.

SiteEvo realizes these two information sources through a fixed system
prompt ($\mathcal{P}_{\rm sys}$) and a dynamically updated search prompt ($\mathcal{P}^{\rm dyn}_{s,r}$). As illustrated in
Fig.~\ref{fig:stage1_prompt}, the agent input at round $r$ is
$
\mathcal{P}_{s,r}
=
\mathcal{P}_{\rm sys}
\oplus
\mathcal{P}^{\rm dyn}_{s,r}
$, where $\oplus$ denotes prompt concatenation.
The fixed prompt $\mathcal{P}_{\rm sys}$ specifies the detector-design
problem and executable contract defined above, while providing
representative communication mechanisms as domain cues. In contrast, the dynamic prompt
$\mathcal{P}^{\rm dyn}_{s,r}$ evolves with the search by combining
site-specific diagnostics and reference performance with evidence from
previously evaluated structures, including their executable programs,
design hypotheses, learned parameters, and BERs.

\begin{figure}[t]
\centering
\footnotesize

\begin{promptbox}{
Fixed system prompt $\mathcal{P}_{\rm sys}$
\hfill{\normalfont\itshape\scriptsize fixed across rounds}
}
\textbf{Task.}
You are an algorithm-discovery agent for uplink massive-MIMO signal
detection.

\vspace{0.4mm}
\textbf{Wireless model.}
$\mathbf y=\mathbf H\mathbf x+\mathbf n$,
$\mathbf H\in\mathbb C^{256\times128}$, QPSK,
128 single-antenna UEs, and a $16\times16$ UPA BS.

\vspace{0.4mm}
\textbf{Detector contract.}
Each candidate defines
\texttt{prepare}$(\mathbf H,\sigma^2,\boldsymbol{\theta})$
and 
\texttt{detect}$(\mathbf z,\mathbf y,\boldsymbol{\theta})$.

\vspace{0.4mm}
\textbf{Structure--parameter division.}
The agent specifies the detector structure and trainable
parameterization; external optimization learns the numerical parameters.

\vspace{0.4mm}
\textbf{Domain cues.}
Residual refinement, interference cancellation, preconditioning, etc.,
are knowledge examples rather than a fixed block library.
\end{promptbox}

\begin{promptbox}{
Dynamic round-$r$ prompt $\mathcal{P}^{\rm dyn}_{s,r}$
\hfill{\normalfont\itshape\scriptsize updated after each round}
}
\begin{tabularx}{\linewidth}{@{}>{\bfseries}l X@{}}
Site evidence &
$\boldsymbol\psi_s$ (\texttt{dataset\_summary}): correlation, diagonal dominance,
and received-power spread. \\[0.6mm]

Performance &
\texttt{LMMSE BER} $=1.483\times10^{-2}$,\quad
\texttt{OAMP-Net BER} $=5.548\times10^{-3}$,\quad
\texttt{current best} $=9.492\times10^{-3}$. \\[0.6mm]

Leader evidence &
\texttt{source program} + hypothesis + mechanism +
learned $\boldsymbol{\theta}$ + measured BER/BER-by-SNR. \\[0.6mm]

Search memory &
recent accepted candidates and recent execution rejections. \\[0.6mm]

Instruction &
Propose structurally distinct detector programs using the measured
evidence and executable leaders; minimize validation BER, with
complexity deferred to Stage~II.
\end{tabularx}
\end{promptbox}
\vspace{-4mm}
\caption{Simplified illustration of the Stage-I prompt. The fixed system
prompt specifies the detector-design problem, executable contract, and
domain knowledge, whereas the dynamic prompt injects site-specific
evidence and accumulated search experience at each search round.}
\label{fig:stage1_prompt}
\vspace{-6mm}
\end{figure}

Conditioned on these two sources of information, the LLM agent proposes
structurally distinct candidates $\mathcal{A}$, each instantiated as a
complete executable detector comprising channel-dependent preparation
and payload detection, implemented by
$\operatorname{prepare}(\cdot)$ and $\operatorname{detect}(\cdot)$,
respectively, together with an associated trainable parameterization.
These candidates are returned in a structured JSON format containing
their executable PyTorch source code, trainable parameter specifications,
and corresponding design rationales.

{\bf (ii) Execution and parameter learning.} Once a candidate structure
$\mathcal{A}$ is generated, SiteEvo first checks whether its executable
program satisfies the common design contract. If execution raises an
exception, the LLM agent is re-invoked with the original candidate
and the observed error message to perform an execution-only repair.
The repair prompt explicitly requires the agent to preserve the
algorithmic hypothesis and computational structure, while withholding
BER, ranking, and complexity information. This isolates code-level
correction from performance-driven detector discovery.

For each executable structure $\mathcal{A}$, its associated parameters
are then learned from the site-specific training data ($\mathcal{D}_{s}^{\rm tr}$) as
\begin{equation}
\widehat{\boldsymbol{\theta}}_{\mathcal{A},s}
=
\argmin_{\boldsymbol{\theta}\in\Theta(\mathcal{A})}
L_{\rm tr}
\left(
\mathcal{A},\boldsymbol{\theta};
\mathcal{D}_{s}^{\rm tr}
\right),
\label{eq:param_learning}
\end{equation}
where $L_{\rm tr}$ denotes the  symbol-level training loss. This
establishes a clear division between structural discovery and numerical
optimization: the LLM agent determines $\mathcal{A}$ and its trainable
parameterization, whereas the backend optimizes the parameter values.

{\bf (iii) BER evaluation and feedback.} After parameter learning, each
candidate is evaluated on the site-specific validation data ($\mathcal{D}_{s}^{\rm val}$) to
characterize its detection BER performance as $
B_s(\mathcal{A})
=
\operatorname{BER}
\left(
\mathcal{A},
\widehat{\boldsymbol{\theta}}_{\mathcal{A},s};
\mathcal{D}_{s}^{\rm val}
\right)$. 
Besides ranking the candidates, this evaluation provides empirical evidence for subsequent discovery. After all
candidates in round $r$ are evaluated, the search archive is updated,
and the current leaders and recent search outcomes are incorporated into
$\mathcal{P}^{\rm dyn}_{s,r+1}$. The next round can therefore reason
from measured relationships among detector structures, learned
parameters, and site-specific BERs. This round-wise feedback enables the
search to progressively refine promising algorithmic directions rather
than repeatedly generating candidates from a fixed detector space.

{\bf (iv) Multi-fidelity selection.} Fully optimizing every generated
candidate during the broad discovery stage would incur excessive learning
cost and unnecessarily limit structural exploration. SiteEvo therefore
uses a modest parameter-optimization budget for initial candidate
evaluation and reserves more intensive optimization for promising
structures. After the discovery rounds, selected candidates are
progressively promoted to higher fidelities with increased optimization
budgets, continuing from their previously learned parameters while
retaining the best validation checkpoint. This provides a more reliable
comparison among competitive structures without imposing the full
training cost on every generated candidate. The detector achieving the
lowest high-fidelity validation BER is finally selected as $(\overline{\mathcal{A}}_{s},
 \overline{\boldsymbol{\theta}}_{s})
$,
and passed to Stage~II as the performance anchor and structural parent.

\begin{figure}[t]
\centering
\footnotesize

\begin{promptbox}{
Fixed Stage-II prompt $\mathcal{P}^{\rm II}_{\rm sys}$
\hfill{\normalfont\itshape\scriptsize fixed across rounds}
}
\textbf{Role.}
You are a structural-refinement agent for an already-discovered
massive-MIMO detector.

\vspace{0.4mm}
\textbf{Parent constraint.}
Every proposal must be a traceable structural descendant of the
supplied Stage-I parent rather than an unrelated detector.

\vspace{0.4mm}
\textbf{Refinement principle.}
Preserve detection performance while reducing block-level computation
through removing, merging, sharing, approximating, or reordering
operations.

\vspace{0.4mm}
\textbf{Parameter inheritance.}
Retained parameters are initialized from Stage-I values; external Adam optimizes all trainable parameters.
\end{promptbox}

\begin{promptbox}{
Dynamic round-$r$ prompt $\mathcal{P}^{\rm II,dyn}_{s,r}$
\hfill{\normalfont\itshape\scriptsize updated after each round}
}
\begin{tabularx}{\linewidth}{@{}>{\bfseries}l X@{}}

Parent evidence &
source program + mechanism + learned $\boldsymbol{\theta}$ +
validation BER + preparation/online/block FLOPs. \\[0.6mm]

Refinement target &
$B_s(\mathcal{A})\le(1+\epsilon)\overline{B}_s$;
minimize
$C_{\rm blk}=C_{\rm pre}+N_{\rm coh}C_{\rm on}$. \\[0.6mm]

Leader evidence & low-cost feasible and low-BER descendants + measured BER and block FLOPs. \\[0.6mm]

Search memory &
recent structural transformations, evaluated outcomes, and execution
rejections. \\[0.6mm]

Instruction &
Propose structurally distinct parent-derived refinements; specify the
structural edits and avoid previously evaluated structures.
\end{tabularx}
\end{promptbox}
 \vspace{-3mm}
\caption{Simplified illustration of the Stage-II prompt. The fixed part enforces communication-aware structural refinement of the Stage-I parent; the dynamic part provides parent evidence, objectives, and prior descendant evaluations.}
\label{fig:stage2_prompt}
\vspace{-4mm}
\end{figure}

\vspace{-3mm}
\subsection{Stage II: Performance-Constrained Structural Refinement}
\label{subsec:stage2}
Stage~I establishes a strong site-specific performance anchor, while
Stage~II seeks a more deployment-efficient realization with limited
performance loss. Unlike the open discovery in Stage~I, Stage~II is
parent conditioned: every candidate must be a traceable refinement of
the selected Stage-I detector. This procedure, depicted in Fig.~\ref{fig:siteevo_overview} (right), is detailed below.

{\bf (i) Parent detector and refinement objective.} Stage~II starts from
the selected parent
$(\overline{\mathcal{A}}_{s},
\overline{\boldsymbol{\theta}}_{s})$ together with its validation BER
and computational profile. Since the channel-dependent operations in
$\operatorname{prepare}(\cdot)$ can be reused across multiple payload
symbols, SiteEvo distinguishes their cost from the computations repeated
during payload detection. Let $C_{\rm pre}(\mathcal{A})$ and
$C_{\rm on}(\mathcal{A})$ denote the corresponding complexities. For
$N_{\rm coh}$ payload symbols sharing the same channel realization, the resulting block-level computational cost is $  C_{\rm blk}(\mathcal{A}) = C_{\rm pre}(\mathcal{A}) + N_{\rm coh}C_{\rm on}(\mathcal{A})$.
Let \(\operatorname{Desc}(\overline{\mathcal A}_s)\) denote the set comprising the Stage-I parent and its traceable structural descendants, and let \(\overline B_s=B_s(\overline{\mathcal A}_s)\) denote the parent's validation BER. Stage~II then seeks the lowest-cost descendant whose BER remains
within a relative tolerance $\epsilon$ of the parent:
\begin{equation}
%\begin{aligned}
\min_{\mathcal{A}\in
\operatorname{Desc}(\overline{\mathcal{A}}_{s})}
~
C_{\rm blk}(\mathcal{A})
~~~\mathrm{s.t.}~~
B_s(\mathcal{A})
\le
(1+\epsilon)\overline{B}_{s}.
%\end{aligned}
\label{eq:stage2_refinement}
\end{equation}

{\bf (ii) Structural descendant generation.} As illustrated in
Fig.~\ref{fig:stage2_prompt}, the refinement agent is driven by a fixed
prompt encoding communication knowledge and parent-descendant constraints,
together with a dynamic prompt containing the parent program, learned
parameters, refinement objective, and previous outcomes. Based on this
context, the LLM reasons about which computations can be modified while
preserving detection performance and proposes traceable transformations,
such as removing, merging, sharing, approximating, or reordering
operations. Thus, refinement is communication-aware rather than blind
module pruning or unrelated replacement.

\begin{table} %[t]
\centering
%\vspace{-5mm}
\caption{Channel configurations of the three deployment sites.}
\vspace{-2mm}
\label{tab:sites}
\scriptsize
\setlength{\tabcolsep}{2.0pt}
\renewcommand{\arraystretch}{0.98}

\begin{tabularx}{\columnwidth}{
@{}
>{\raggedright\arraybackslash}p{0.185\columnwidth}
>{\raggedright\arraybackslash}X
@{}}
\toprule
Site/model & Channel generation and key parameters \\
\midrule

\textbf{A}: Clustered geometric multipath &
32 groups ($8\!\times\!4$; 4 UEs/group); az./el. spans
$[-60^\circ,60^\circ]/[-20^\circ,20^\circ]$; group spreads
$4.5^\circ/2.5^\circ$; $L=4$ paths with spreads $3^\circ/2^\circ$
and powers $[0,-2.5,-5.5,-8]$ dB. \\

\textbf{B}: Double-scattering &
$\mathbf H=\mathbf G_1\mathbf G_2/\sqrt{80}$,
$\mathbf G_1\!\in\!\mathbb C^{256\times 80}$,
$\mathbf G_2\!\in\!\mathbb C^{80\times128}$, with i.i.d.
$\mathcal{CN}(0,1)$ entries; $\operatorname{rank}(\mathbf H)\leq80$. \\

\textbf{C}: Kronecker-correlated Rayleigh &
Exponential Kronecker correlation:
$[{\mathbf R}_{\rm h}]_{i,j}=0.88^{|i-j|}$,
$[{\mathbf R}_{\rm v}]_{i,j}=0.72^{|i-j|}$, and
$[{\mathbf R}_{\rm u}]_{i,j}=0.75^{|i-j|}$ for horizontal, vertical, and
inter-user correlation. \\

\bottomrule
\end{tabularx}
\vspace{-5mm}
\end{table}

{\bf (iii) Validity check, parameter adaptation, and evaluation.}
Each descendant first undergoes the same executable-validity check as
in Stage~I, with execution-only repair applied when necessary. 
Retained parameters are initialized from their learned Stage-I values, while newly introduced parameters are initialized separately. All trainable parameters are then optimized on $\mathcal{D}_{s}^{\rm tr}$.
The adapted detector is then evaluated in terms of validation BER and block-level cost $C_{\rm blk}$. These measurements are fed back to guide subsequent
refinement.

{\bf (iv) Multi-fidelity refinement and selection.} As in Stage~I,
promising descendants are progressively promoted to higher
parameter-optimization budgets for reliable comparison. SiteEvo selects
the feasible descendant with the minimum block-level cost. If none
improves upon the parent while satisfying the BER constraint, the
Stage-I parent is retained. 
The resulting $(\mathcal{A}_{s}^{\rm dep},\boldsymbol{\theta}_{s}^{\rm dep})$ is frozen for deployment, requiring neither LLM inference nor parameter learning online.

\vspace{-5mm}
\section{Numerical Results and Discussion}

\subsection{Simulation Setup}
\label{subsec:simulation_setup}

The uplink massive MIMO system employs a $16\times16$ uniform planar
array (UPA) with $N=256$ antennas at the BS and serves $K=128$
single-antenna UEs using unit-power quadrature phase-shift keying (QPSK) signaling. As summarized in
Table~\ref{tab:sites}, three contrasting deployment conditions are
considered. Site~A represents UEs concentrated in angular hotspots,
leading to strong inter-user interference; Site~B represents a
scatterer-limited environment with restricted spatial degrees of
freedom; and Site~C represents pronounced spatial and inter-user
correlation. This enables evaluation of how
site-dependent propagation characteristics affect detector structure.

A common large-scale fading model and global normalization are applied
across the three sites so that their differences primarily arise from
the underlying channel structures. The normalization preserves relative
user powers as well as the correlation and rank characteristics of each
site. For each site, 2048/256/1024 channel samples are used
for training, validation, and testing, respectively. Training uses signal-to-noise ratios (SNRs) of $\{0,6,12\}$ dB, while
testing uses $\{0,3,6,9,12\}$ dB. The standard test uses 128 payload symbols per channel realization at each SNR. 
Validation data are used for detector search and selection, whereas the
test set is reserved for final evaluation.
The training loss combines symbol mean square error and a QPSK classification loss.

We employ DeepSeek-V4.1-Flash as the LLM agent
\cite{DeepSeekAI2026Flash}. Stage~I starts with two
executable linear minimum mean-square error (LMMSE)-based residual-correction detectors
and performs 45 search rounds, with up to three
proposals per round. Each generated detector is
limited to 64 trainable scalar parameters.
Candidates are initially trained for 150 Adam optimizer steps, with promising candidates progressively promoted to total budgets of 600, 3000, and 6000 optimization steps. Stage~II performs 20 rounds of parent-conditioned
structural refinement with up to three proposals per round. We set
$N_{\rm coh}=200$ and the BER-preservation tolerance to $\epsilon=0.1$. The DeepSeek API costs for Sites A, B, and C were RMB 5.89, 3.97, and 5.38, respectively.

\vspace{-3mm}

\subsection{Detection Performance and Structural Insights}
\label{subsec:performance}

We compare SiteEvo with LMMSE, site-specific OAMP-Net, and two implemented structure-search baselines, LLM-DSL and Evolution. For OAMP-Net, $T\in\{8,12,16,20\}$ is selected by site-wise validation BER. Both structure-search baselines use the same predefined grammar with up to five stages, each selected from residual, soft-decision, momentum-assisted, and diagonally gated corrections. LLM-DSL employs LLM-guided structure search, whereas Evolution uses an evolutionary search strategy. 
All three structure-search methods use a matched budget of 135 candidate evaluations and a common parameter-optimization protocol.

 \begin{figure}[t]
    \centering
    \includegraphics[width=0.95\linewidth]{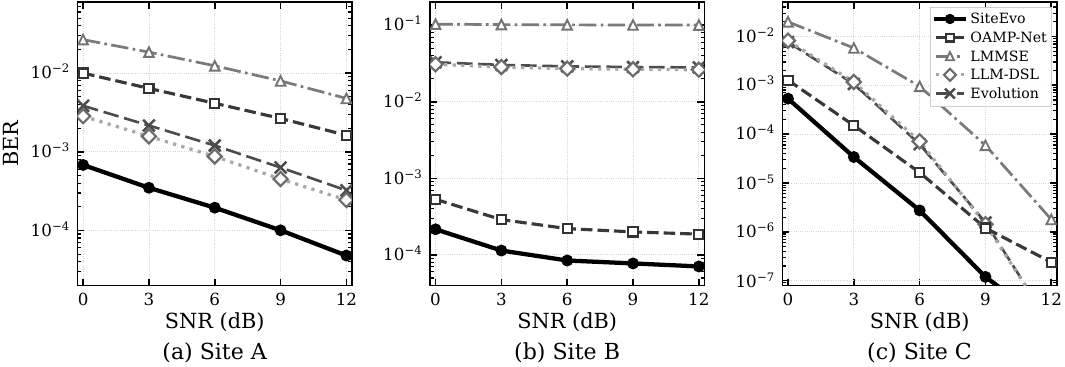}
    \vspace{-3mm}
     \caption{Stage-I BER versus SNR for (a) clustered geometric,
    (b) rank-deficient double-scattering, and
    (c) Kronecker-correlated Rayleigh channels.}
    \label{BERresults}
    \vspace{-5mm}
\end{figure}

\subsubsection{Detection performance}
Fig.~\ref{BERresults} compares the Stage-I detectors across the three channel environments. 
SiteEvo achieves the lowest measured BER across all three channel environments and evaluated SNRs.
The relative performance of competing methods changes substantially across sites. At Site~A, both LLM-DSL
and Evolution outperform OAMP-Net, whereas at Site~B, these two methods exhibit higher BER than OAMP-Net and
only limited improvement with increasing SNR.
In contrast, SiteEvo maintains its advantage across all three channel distributions, reducing the BER
of the strongest competing detector at 6~dB by factors of approximately 4.5, 2.6, and 6.0 at Sites A, B,
and C, respectively. 
While BER decreases markedly
with SNR at Sites A and C, the curves at Site~B gradually flatten, reflecting
detection challenges imposed by its rank-deficient channel.

\subsubsection{Site-adaptive structural discovery analysis}

At Site~A, Stage~I discovers a two-pass detector initialized by a regularized linear estimate.
Each pass combines learned residual updates, magnitude-dependent gating, momentum, residual feedback correction, and damped soft QPSK projection.
The first pass performs 12 residual updates followed by a 12-step Chebyshev-type polynomial correction,
while the second performs six updates followed by a six-step correction. Both polynomial stages
use a $4\times4$ block-Jacobi preconditioner
\begin{equation}
\mathbf P_{\rm A}=
\operatorname{blkdiag}_{g=1}^{32}
\left[
(\mathbf W_{\mathcal I_g,\mathcal I_g}
+\lambda\sigma^2\mathbf I_4)^{-1}
\right],
\label{eq:siteA_preconditioner}
\end{equation}
where $\mathbf W=\mathbf H^H\mathbf H$,
$\mathcal I_g=\{4g-3,\ldots,4g\}$,
and $\lambda$ is a learned regularization parameter.
The polynomial correction directions follow a Chebyshev-type recurrence involving
$\mathbf P_{\rm A}\mathbf W$.
The $4\times4$ blocks match Site~A's grouped user ordering, although group labels are not provided to the agent.

At Site~B, Stage~I discovers a detector combining 11 Cholesky-based residual-update iterations with rank-12 spectral corrections constructed
through deflated power iterations.
Each iteration applies residual-gated spectral correction and damped soft QPSK projection.
Four subsequent refinement steps reuse the Cholesky and spectral corrections, followed by a final damped QPSK-aware soft projection to improve detection under rank deficiency.
At Site~C, Stage~I constructs three Cholesky-based preconditioners by deflating the dominant 4, 8, and 12 eigenmodes of the diagonally normalized
Gram matrix. Fourteen residual-update iterations combine their corrections with 2-step momentum and soft QPSK projection. The detector then
performs 3 additional preconditioned residual
updates and 2 successive passes of pairwise block correction.

\subsubsection{Complexity and cross-site adaptation}
As shown in Table~\ref{tab:complexity}, Stage~II
reduces the Stage-I block-level floating-point operations (FLOPs) by 16.7\%, 27.9\%, and 34.1\%
at Sites A, B, and C, respectively, while satisfying the validation-BER constraint.
At Site~C, for example, refinement reduces 14 three-preconditioner iterations to eight, followed by four iterations using a single preconditioner, while eliminating three
additional residual updates and one of the two pairwise block-correction passes. At 3~dB, the refined detectors achieve BERs of $(3.69, 0.95, 0.34)\times10^{-4}$, representing a 16.7\% reduction at Site~B, a 5.8\% increase at Site~A, and nearly unchanged BER at Site~C. Compared with OAMP-Net, the refined detectors require 52.3--63.1\% fewer block-level FLOPs.

Table~\ref{tab:transfer} evaluates cross-site detector structure transfer at 3~dB after target-site parameter finetuning. All transferred detectors exhibit higher BERs than their site-specific counterparts. For A-to-B transfer, the BER remains approximately 50\% higher than that of the native Site-B detector. For B-to-C transfer, the BER reaches $5.28\times10^{-4}$, approximately 15.4 times the native Site-C BER. These results demonstrate the limitations of parameter adaptation alone and highlight the importance of site-specific detector structure discovery.

\begin{table}[!t]
\centering

% ===== Table II: BER at 3 dB and FLOPs =====
\begin{minipage}[t]{0.55\columnwidth}
\vspace{0pt}
\centering
\caption{Test BER at 3 dB ($10^{-4}$) /
200-symbol FLOPs ($10^9$).}
\label{tab:complexity}
\vspace{-2mm}
\scriptsize
\setlength{\tabcolsep}{1pt}
\renewcommand{\arraystretch}{1.00}

\begin{tabular*}{\linewidth}{
@{\extracolsep{\fill}}lccc@{}
}
\toprule
Site\vphantom{
\diagbox[
width=7.2em,
height=4.2ex,
innerleftsep=1pt,
innerrightsep=1pt
]{Source}{Target}
}
& OAMP-Net & Stage I & Stage II \\
\midrule
A & 64.12/2.59 & 3.49/1.15 & 3.69/0.95 \\
B & 2.89/1.96 & 1.14/1.12 & 0.95/0.81 \\
C & 1.50/2.59 & 0.34/1.87 & 0.34/1.23 \\
\bottomrule
\end{tabular*}
\end{minipage}
\hfill
% ===== Table III: Cross-site transfer at 3 dB =====
\begin{minipage}[t]{0.42\columnwidth}
\vspace{0pt}
\centering
\caption{Cross-site BER at 3 dB after
fine-tuning ($10^{-4}$).}
\label{tab:transfer}
\vspace{-2mm}
\scriptsize
\setlength{\tabcolsep}{1pt}
\renewcommand{\arraystretch}{1.00}

\begin{tabular*}{\linewidth}{
@{\extracolsep{\fill}}lccc@{}
}
\toprule
\diagbox[
width=5em,
height=4.2ex,
innerleftsep=1pt,
innerrightsep=1pt
]{Source}{Target}
& A & B & C \\
\midrule
A & 3.69 & 1.42 & 7.81 \\
B & 57.56 & 0.95 & 5.28 \\
C & 28.43 & 1.78 & 0.34 \\
\bottomrule
\end{tabular*}
\end{minipage}
\vspace{-5mm}
\end{table}

\vspace{-3mm}
\section{Conclusion}
This letter proposed SiteEvo, an agentic framework
for site-adaptive massive MIMO detector discovery.
Guided by pretrained communication knowledge,
site-specific channel evidence, and search feedback,
the LLM agent generates executable detector structures
and trainable parameterizations. A communication
backend optimizes their parameters and returns
measured BER to guide further discovery.
A second agentic stage reduces computational cost
under a BER-preservation constraint. The resulting
detectors are frozen for deployment without online
LLM inference. Simulations across three sites demonstrate distinct
site-specialized structures with improved detection performance and computational efficiency.

\vspace{-3mm}
 
 \bibliographystyle{IEEEtran}
\bibliography{reference}

\end{document}